\documentclass[manuscript, screen]{acmart}

\usepackage{graphicx}
\usepackage{subfig}
\usepackage{tabularray}
\usepackage{booktabs}
\usepackage{multirow}
\usepackage{array}
\usepackage{tabularx}
\usepackage{enumitem}
\usepackage{colortbl}
\usepackage{xcolor}
\usepackage{amsmath}
\definecolor{mygray}{RGB}{204,204,204}
\newcommand{\grayline}{\arrayrulecolor{mygray}\hline}
\newcolumntype{P}[1]{>{\centering\arraybackslash}p{#1}}
\AtBeginDocument{%
  }

\copyrightyear{2024}
\acmYear{2024}
\setcopyright{rightsretained}
\acmConference[LAK '24]{The 14th Learning Analytics and Knowledge Conference}{March 18--22, 2024}{Kyoto, Japan}
\acmBooktitle{The 14th Learning Analytics and Knowledge Conference (LAK '24), March 18--22, 2024, Kyoto, Japan}\acmDOI{10.1145/3636555.3636896}
\acmISBN{979-8-4007-1618-8/24/03}

\begin{document}

\title[Improving Learning with Hybrid Human-AI Tutoring]{Improving Student Learning with Hybrid Human-AI Tutoring: A Three-Study Quasi-Experimental Investigation }

\author{Danielle R. Thomas}
\email{drthomas@cmu.edu}
\affiliation{
  \institution{Carnegie Mellon University}
  \city{Pittsburgh}
  \country{USA}
}

\author{Jionghao Lin}
\email{jionghaol@andrew.cmu.edu}
\affiliation{%
  \institution{Carnegie Mellon University}
  \city{Pittsburgh}
  \country{USA}
}

\author{Erin Gatz}
\email{egatz@andrew.cmu.edu}
\affiliation{%
  \institution{Carnegie Mellon University}
  \city{Pittsburgh}
  \country{USA}
  }

\author{Ashish Gurung}
\email{agurung@andrew.cmu.edu}
\affiliation{%
  \institution{Carnegie Mellon University}
  \city{Pittsburgh}
  \country{USA}
  }

\author{Shivang Gupta}
\email{shivang@cmu.edu}
\affiliation{%
  \institution{Carnegie Mellon University}
  \city{Pittsburgh}
  \country{USA}
  }

\author{Kole Norberg}
\email{Iknorberg1@carnegielearning.com}
\affiliation{%
  \institution{Carnegie Learning, Inc.}
  \city{Pittsburgh}
  \country{USA}
  }

\author{Stephen E. Fancsali}
\email{sfancsali@carnegielearning.com}
\affiliation{%
  \institution{Carnegie Learning, Inc.}
  \city{Pittsburgh}
  \country{USA}
  }

\author{Vincent Aleven}
\email{aleven@cs.cmu.edu}
\affiliation{%
  \institution{Carnegie Mellon University}
  \city{Pittsburgh}
  \country{USA}
  }

\author{Lee Branstetter}
\email{branstet@andrew.cmu.edu}
\affiliation{%
  \institution{Carnegie Mellon University}
  \city{Pittsburgh}
  \country{USA}
  }

\author{Emma Brunskill}
\email{ebrun@cs.stanford.edu}
\affiliation{%
  \institution{Stanford University}
  \city{Stanford}
  \country{USA}
  }

\author{Kenneth R. Koedinger}
\email{koedinger@cmu.edu}
\affiliation{
  \institution{Carnegie Mellon University}
  \city{Pittsburgh}
  \country{USA}
}


\renewcommand{\shortauthors}{Thomas et al.}

\begin{abstract}

Artificial intelligence (AI) applications to support human tutoring have potential to significantly improve learning outcomes, but engagement issues persist, especially among students from low-income backgrounds. We introduce an AI-assisted tutoring model that combines human and AI tutoring and hypothesize this synergy will have positive impacts on learning processes. To investigate this hypothesis, we conduct a three-study quasi-experiment across three urban and low-income middle schools: 1) 125 students in a Pennsylvania school; 2) 385 students (50\% Latinx) in a California school, and 3) 75 students (100\% Black) in a Pennsylvania charter school, all implementing analogous tutoring models. We compare learning analytics of students engaged in human-AI tutoring compared to students using math software only. We find human-AI tutoring has positive effects, particularly in student’s proficiency and usage, with evidence suggesting lower achieving students may benefit more compared to higher achieving students. We illustrate the use of quasi-experimental methods adapted to the particulars of different schools and data-availability contexts so as to achieve the rapid data-driven iteration needed to guide an inspired creation into effective innovation. Future work focuses on improving the tutor dashboard and optimizing tutor-student ratios, while maintaining annual costs per student of approximately \$700 annually.

\end{abstract}




\maketitle

\section{Introduction}

The main obstacle to improving math performance among middle school students lies in ensuring fair and equal access to effective learning opportunities \cite{chine2022educational, kraft2021blueprint}. While economically disadvantaged and historically underserved students have the potential to excel when given the same resources as their peers (c.f., \cite{koedinger2023astonishing}), they often face learning gaps due to limited access \cite{chine2022educational}. Individualized instruction via tutoring can have consistent and significant positive impacts on student achievement and learning \cite{guryan2021not, kraft2021blueprint}, particularly when deployed among middle and high school grades in math and during the school day, as opposed to after school \cite{nickow2020impressive} and when delivered by trained tutors attending to students' socio-motivational needs and relationship building \cite{chhabra2022evaluation, kraft2021school}. However, low-income students lack access to in-school tutoring programs and well-trained tutors, evidenced by the 16 million low-income children on the waitlist for high-quality afterschool programs \cite{afterschool2020}. Further exacerbating the opportunity gap among students from low-income families is the high costs of \$2500+ per student for private, individual tutoring \cite{kraft2021school}—most families simply cannot afford it. Relying solely on human tutoring cannot adequately address the current educational needs, which have intensified in the face of pandemic-era learning losses \cite{west2023edtech}. The gravity of the situation is evidenced by declines in middle school math scores reported in 2022 – the steepest since the inception of state assessments \cite{usdoe2022naep}. The lack of accessibility faced by students arises from a range of factors, such as inadequate access to basic inputs like digital devices and internet connectivity. Additionally, there are complications related to inclusivity, which inadequately addresses the diverse needs of students, including English language learners and those with disabilities \cite{west2023edtech}. The challenges facing math learning related to access, equity, fairness, and inclusion have fostered collaborative and focused efforts on AI-assisted human-technology ecosystems that increase learning opportunities for all students  \cite{chine2022educational, thomas2023ai}.

High-dosage human tutoring has been shown to double student learning \cite{nickow2020impressive, pane2014effectiveness}. However, providing intensive, individualized \textit{human} tutoring to the millions of students deemed in need of support is costly and requires more human capital than society is likely to mobilize \cite{kraft2021blueprint}—hence, the emergence of  AI as a complement to human tutoring. Advances in AI and learning analytics have created opportunities to increase tutoring efficiency and lower tutor-to-student ratios \cite{aleven2023towards, chine2022educational}. AI-assisted (also known as, AI-augmented, AI-in-the-loop, AI-supported) human tutoring works by having the AI provide the adaptive math instruction, heavily researched and well-known to be effective \cite{nickow2020impressive}, with humans providing as-needed socio-motivational intervention and relationship-building support \cite{chine2022educational}. An example in practice is the application of a real-time AI-driven dashboard to support effective use of tutor's time during tutoring and differentiating student support by allocating tutor time to students who need it most. There is evidence of this approach among teachers demonstrating students learn more when teachers and AI work in synergy \cite{holstein2022designing}. This study among teachers is the only one known of its kind and sparks the question- what about human tutors? In \cite{chine2022educational}, among 70 participating students from predominantly low-income backgrounds in a human-AI synergistic tutoring program, an observed doubling of learning was demonstrated compared to a matched control group. However, this prior study took place within an after-school program that spanned the COVID pandemic.

\begin{figure*}[h]
\centering
  \includegraphics[width=0.9\textwidth]{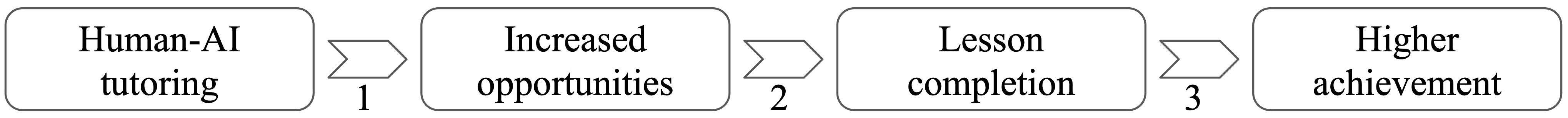}
\vspace{3mm}
\caption{Conceptual pathway connecting human-AI tutoring to higher student achievement by increasing learning opportunities (1), yielding more lesson completions (2), and subsequently leading to higher achievement (3).}

\label{fig:intro_1}

\end{figure*}

This current work focuses on this novel AI-informed approach to tutoring and occurs during the school day. Informally referred to as \textit{hybrid human-AI tutoring} and described as human tutors and AI-based tutoring software working in tandem, this method provides just-in-time assistance to students and is scalable, by allowing a single tutor to work with more students. Hybrid human-AI tutoring shows promise as a means of doubling student learning at a fraction of the cost of high-dosage human tutoring \cite{aleven2023towards, chine2022educational}. However, its impact on student learning processes and intermediate outcomes leading to improving achievement remains presently unknown. Fig. \ref{fig:intro_1} provides a conceptual pathway connecting participation in human-AI tutoring to higher achievement, shown in the following series of steps: 1) student participation in human-AI tutoring increases exposure to learning opportunities; 2) increases in learning opportunities yields more successful lesson (i.e., module, unit) completions; and 3) more lesson completions yields higher achievement. Research supports that students of all ability levels, ranging from those several grade levels behind to accelerated learners, can benefit from tutoring \cite{koedinger2023astonishing}. However, there is currently a lack of research on how hybrid human-AI tutoring benefits students across varying baseline abilities. This present tutoring approach is specifically designed to attend to students from low-income backgrounds, constituting over 90 percent of students in the study. 
\vspace{3mm}

One key challenge is that the assessment of effectiveness remains a protracted endeavor. Randomized controlled trials (RCTs), are universally accepted as the ideal methodology for assessing intervention efficacy and impact \cite{miller2020experimental}. However, conducting RCTs necessitates a substantial investment of time, human capital, and resources, often spanning across multiple years to yield conclusive results \cite{hariton2018randomised}. Furthermore, the implementation of RCTs raises ethical quandaries, requiring certain students to be arbitrarily denied beneficial interventions that, under ordinary circumstances, are most likely superior to no intervention at all. Quasi-experimental studies may be underutilized as a means of evaluating early program effectiveness, particularly when RCTs are expensive and take a long time to complete. In addition, RCTs alone are not a ``holy grail'' of effectiveness research, as they often provide insight into ``what works'' but not ``why it works,'' with other methods often needed to pinpoint the causal mechanisms \cite[p.~2]{deaton2018understanding}. As AI-supported tutoring programs are a new approach, there simply is not enough empirical research published leveraging RCTs to gauge early program impact on student learning. While rapid quasi-experimental studies (e.g., pre-post with non-equivalent control groups) offer a more agile and resource-efficient means of assessing intervention outcomes, they are not without limitations, such as the lack of random assignment introducing potential biases and confounding factors that may impact internal validity \cite{miller2020experimental}. Strategies to mitigate biases, such as matching techniques, can remove some potential bias providing useful, rapid, and relatively low-cost analysis of  impact \cite{chine2022educational, miller2020experimental}.

Given this, we focus on enhancing the efficiency and rigor of experimental trials of the hybrid human-AI tutoring approach by conducting rapid-cycle quasi-experimental studies, which can streamline and expedite the evaluation of program effectiveness. The implementation of several rapid-cycle studies across different school sites and student populations can create multiple lines of evidence which combine to generate a more comprehensive evaluation of efficacy. Our three-study quasi-experimental investigation aims to determine the initial impact of a hybrid human-AI tutoring program that takes place during the regular school day and focuses on low-income students. In providing hybrid AI-human tutoring support to students who need it most within three ethnically and geographically diverse U.S. middle schools, we strive to answer the following research questions:

\begin{itemize}[left=0pt, label={}]
    \item[] \textbf{RQ 1:} What differences exist in learning processes and intermediate outcomes between students participating in hybrid human-AI tutoring compared to students engaging solely with math software? 

    \item[] \textbf{RQ 2:} Does the impact of hybrid human-AI tutoring vary by student's baseline ability? Specifically, if there are positive effects on student learning processes, is the effect equally distributed across students of varying baseline abilities? 
\end{itemize}

\section{Related Work}

\subsection{Human-delivered High-Impact Tutoring} High-impact tutoring is a targeted and intensive form of academic support designed to provide personalized assistance to students who may be struggling academically or require additional help beyond the capabilities of regular classroom instruction \cite{accelerator2021toolkit}. It is characterized by several key features, including: provision during the school day three or more times per week; delivery by well-trained tutors; usage of student data to guide instruction; and limitation of teacher-to-student ratios of 1-to-3 or lower \cite{accelerator2021toolkit}. Across eight meta-analyses (see e.g. \cite{nickow2020impressive}) encompassing over 150 studies, tutoring has shown to substantially improve student learning, especially among underrepresented and marginalized students, though the effect size can vary with different program characteristics. In a RCT consisting of 2,600 9th- and 10th-grade students attending Chicago Public Schools and using Saga Education's high-dosage tutoring model, researchers found tutoring increased math test scores by 0.16 standard deviations (SD), with later replications reporting 0.37 SD gains \cite{guryan2021not}. Not surprisingly, given the substantial investment in human capital and the time skilled tutors spend with students, high-dosage tutoring is costly, with Saga's model averaging \$3,500 to \$4,300 per student annually \cite{guryan2021not}.  

\subsection{Impact of AI-assisted Math Software} Given the high cost of high-impact tutoring delivered exclusively by humans, researchers have long explored the less expensive alternative of creating intelligent tutoring systems (ITS) and related math software to personalize math instruction and boost learning for K-12 students. Specifically, AI-based technologies has been shown to facilitate the math software systems across various dimensions, including creating rich environments (e.g., multimedia applications), fostering individual functionality of multiple components (e.g., learner model, domain model, pedagogical model) within modular architecture and their communication, and delivering adaptive instructions grounded in educational and psychological principles \cite{corbett2001cognitive}. The overarching impact of AI on mathematical learning involves almost all scales ranging from ``physical intelligence'' to ``digital intelligence.'' A large number of review studies and meta-analyses suggest large learning gains can be achieved when students use these technologies \cite{escueta2020upgrading}, but there can be substantial variability in usage of educational technology, and such variability in usage can be correlated with differing learning outcomes. This view receives strong support from important new evidence based on 1.3 million observations of student usage of learning software (including intelligent tutoring systems) across multiple grade levels and subjects points to remarkably similar rates of student learning as a function of practice opportunities, despite large differences in the students’ family income levels, ethnicity, and prior measured academic achievement. These results suggest that educational achievement gaps are driven by differences in learning opportunities \cite{koedinger2023astonishing}. Intelligent tutors show great promise in terms of providing appropriately calibrated learning opportunities to large numbers of students in a cost effective way, but for this promise to be realized students must actually use the technology. A cascade of enabling conditions, including physical and emotional health, access to technology, a growth mindset, and adequate motivation, must be in place; otherwise students will not use the technology, even when it is low-cost and increasingly widely available. We seek to expand access to these promising technologies by using a hybrid human-AI tutoring model, in which we can increase engagement and use of AI math software by the students who need it most.

\subsection{Defining Hybrid Human-AI Tutoring} While intelligent tutoring systems can support tutoring at scale, they are not equipped to provide the relational and motivational support some students may need to engage with these systems \cite{chine2022educational}. Recently, \textit{hybrid human-AI tutoring}, defined, as mentioned, as human and AI tutors working in tandem, aims to leverage the power of AI-driven adaptive math software to provide personalized instruction for students \textit{and} to allow human teachers and tutors to focus more of their effort on the relationship building and socio-emotional support many students need to successfully engage with the AI \cite{heffernan2014assistments, holstein2017intelligent}. Moreover humans can provide additional content support beyond the capability of the AI software system, such as supporting students who are unproductively struggling \cite{holstein2018student}. In recent research and educational practice, human-AI systems have taken on a range of forms. \cite{chine2022educational} and \cite{muralidharan2019disrupting} have examined \textit{schedule-driven} systems in which students are connected with a human tutor through pre-scheduled lessons, often occurring outside regular math classes and featuring a long period of repeated interactions between a student and the same human tutor. Schedule-driven tutoring can be effective, but the costs are high, limiting access and scalability. At the other extreme are \textit{student-driven} systems, in which the students themselves determine whether and what sort of help they need and initiate engagement. This approach can have low cost, but many students who may benefit will not seek support on their own, and the efficacy of these systems for the least advantaged students remains unproven \cite{aleven2016instruction}.  

\begin{figure*}[h]
\centering
  \includegraphics[width=0.9\textwidth]{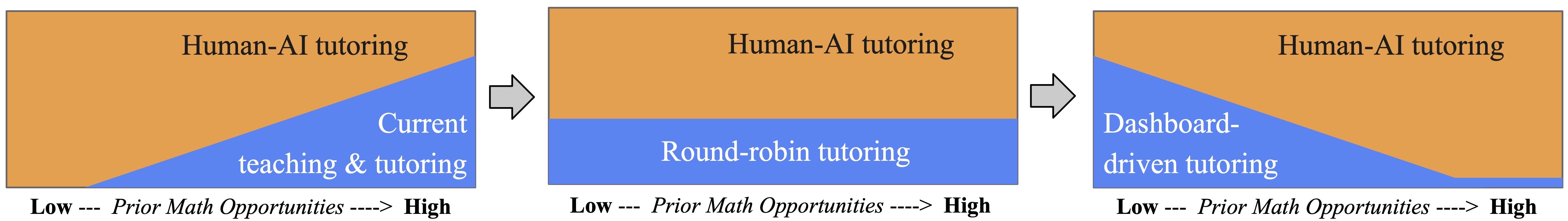}
\caption{Tutor- and dashboard-driven human-AI tutoring can redirect teacher/tutor effort to support the students with the lowest level of prior math opportunities.}

\label{fig:intro_2}
\vspace{-2mm}
\end{figure*}

Tutor-driven and dashboard-guided systems represent a middle ground, in which students spend part of their regular math class time engaged in personalized practice using math software, enhanced with algorithms that can identify student learning progress and struggle that is likely to be unproductive, which are then displayed on a dashboard to inform teachers or tutors, but allow the teachers/tutors to determine which students to support. Alternatively, the dashboard could suggest, in real time, that the classroom teacher and/or virtual tutors support specific struggling students. Either of these systems is likely to be less costly than schedule-driven systems, but more likely than student-facilitated systems to support the learning of the least advantaged and least confident math learners. \citeauthor{holstein2018student} \cite{holstein2018student} showed how this kind of system for classroom teachers tripled students' learning gains in a short-term research study. The system measurably redirected  the teacher's attention towards the students who had lower initial knowledge\textemdash and away from the students most likely to seek assistance (who are often the more confident and adept math learners). Fig. \ref{fig:intro_2} distinguishes between (a) a scenario in which the initiative taken by more confident students deflects teacher/tutor attention away from the students who need attention the most; (b) a scenario in which a ``round robin'' approach tries to ensure an equal distribution of teacher/tutor attention, and (c) our target scenario, in which an intelligent dashboard directs human teacher/tutor attention to the students with the lowest levels of initial math preparation, even when these students are not actively seeking assistance.

We seek to study tutor- and dashboard-driven human-AI tutoring systems in which off-site tutors can initiate remote sessions with students through teleconferencing software during the regular school day.  The availability of these off-site tutors could, by itself, improve student outcomes even if dashboards neither provided much useful information to the tutors on student learning challenges nor directed them to support particular students. Our research design therefore includes AI-human tutoring interventions with and without ``dashboards,'' in order to help us better understand the impact generated by use of a dashboard that provides actionable information and direction.

\section{Methods}

We piloted a hybrid human-AI tutoring model in three U.S. middle schools located in the Midwest, the East, and the West respectively, with students receiving the intervention one day per week during the school day. Denoted as Sites 1 to 3, students interacted with the following math software, respectively: IXL, a comprehensive K-12 math curriculum; i-Ready, a K-8 learning program created by Curriculum Associates; and Carnegie Learning's MATHia (formerly Cognitive Tutor, \cite{ritter2007cognitive}), an  adaptive one-on-one math learning platform designed for grades 6 to 12.\footnote{\url{https://www.ixl.com/}; \url{https://www.curriculumassociates.com/programs/i-ready-learning}; \url{https://www.carnegielearning.com/solutions/math/mathia/}} For Sites 1 and 2 (tutor-driven), weekly student learning process data was collected to determine student's individual needs, but tutors used a “round robin” approach, ensuring they provided support to every student. For Site 3 (dashboard-driven), real-time learning process and usage log data was collected using MATHia's LiveLab, a live tutor-facing dashboard that provides tutors with real-time learning process data, such as idle time and progress, to assist tutors with prioritizing and differentiating student support.\footnote{\url{https://support.carnegielearning.com/help-center/math/livelab/article/livelab-overview/}} Table \ref{tab:met_1} summarizes the implementation characteristics across sites.

\begin{table}[h]
\centering
\small
\caption{Summary of implementation characteristics across sites.}
\vspace{-0.25cm}
\label{tab:met_1}
\resizebox{0.9\textwidth}{!}{%
\renewcommand{\arraystretch}{1.2}
\begin{tabular}{p{4cm} p{2.5cm} p{2.5cm} p{2.5cm}}
\toprule
& \textbf{Site 1} & \textbf{Site 2} & \textbf{Site 3} \\
\hline
EdTech & IXL & i-Ready & MATHia \\
Tutor:student ratio & 1:4 & 1:4 & 1:8 then 1:4 \\
Dashboard & No & No & Prototype \\
Tutoring Strategy & Round robin & Round robin & Dashboard-driven \\
\bottomrule
\end{tabular}
}
\end{table}

\subsection{Site 1 Method}
Site 1 took place in an urban Pennsylvania school district at a middle school enrolling 490 students, of which 99\% were economically disadvantaged and 96\% score below math proficiency. Student demographics were 73\% Latinx, 18\% Black, 3\% White, 1\% Asian, and 5\% multiracial. The entire 7th grade population, excepting nine students due to data loss, (n=125) were arbitrarily assigned by school administrators to either a Control (n=73) or a deferred treatment, in which students received the same condition as the Control initially and then participated in the Delayed Treatment (n=52).

During fall semester, defined as August 22, 2022 through January 18, 2023 (16 weeks), Control and Delayed Treatment groups rotated through special classes with \textit{supplemental math support} defined as students using IXL math software with a non-math (i.e., social studies, science, English/language arts) teacher facilitating the session but providing no further guided math support (\textit{Edtech\_Only}). Beginning in early spring, January 19, 2023 through April 6, 2023 (6 weeks), the Delayed Treatment students received supplemental math support from their math teacher, who provided guided help while they engaged in IXL (\textit{EdTech$+$MathTeacher}). In late spring, April 7, 2023 through May 22, 2023 (7 weeks), the Delayed Treatment students continued to receive supplemental math support from their math teacher, but they were also assisted by tutors initiating remote sessions with individual students through teleconferencing software during support time (\textit{EdTech$+$MathTeacher$+$Tutoring}). Tutor-to-student ratios averaged 1:4. Table \ref{tab:met_2} summarizes the interrupted time series design comparing the Control group to the three conditions within the Delayed Treatment group: math software use only; math software use with a math teacher; and math software use with a math teacher and tutors.

\begin{table*}[htb]
\centering
\small
\caption{The Control used EdTech throughout (\textit{EdTech\_Only}). The Delayed Treatment group added math teacher supervision (\textit{EdTech+MathTeacher}) in Early Spring and remote human tutoring (\textit{EdTech+MathTeacher+Tutoring}) in the Late Spring.}
\resizebox{0.9\textwidth}{!}{%
\begin{tabular}{llll}
\toprule
& \textbf{Fall} & \textbf{Early Spring} & \textbf{Late Spring} \\
\midrule
Control & EdTech\_Only & EdTech\_Only & EdTech\_Only \\
Delayed Treatment & EdTech\_Only & EdTech+MathTeacher & EdTech+MathTeacher+Tutoring \\
\bottomrule
\end{tabular}
}
\label{tab:met_2}
\end{table*}

Throughout the school year, IXL-reported process data regarding skill usage and proficiency was collected weekly and made available to teachers. In addition, Renaissance Learning's Star diagnostic assessment was administered several times to assess student’s baseline comprehension and estimated grade level performance.

\subsection{Site 2 Method} 
Site 2 occurred at a large, urban California school district (n=15,700 students) in one of three middle schools. At the school where Site 2 was conducted, the 385 7th grade student participants were 90\% low-income, 80\% Black or Latinx, and 49\% female. For each student, i-Ready learning process data recorded \textit{time on task} and \textit{time spent completing a lesson}, which were viewed as step 1 measures, and \textit{lessons passed}, which were viewed as a step 2 measure quantifying higher achievement (See Fig. \ref{fig:intro_1}). Students engaged in \textit{i-Ready Personalized Instruction}, as a weekly math intervention, which took place one day per week for 50 minutes, and completed the \textit{i-Ready Diagnostic}, a computer-delivered diagnostic assessment \cite{curriculumassociates2023}. Once students completed the \textit{i-Ready  Diagnostic}, which was administered in the fall 2022, a personalized lesson plan was developed based on individual student's needs, though sometimes teachers also assigned specific lessons separate from the diagnostic-suggested lessons. Then, the last three weeks of the school year (weeks of May 15th, May 22nd, and May 29th, 2023), students engaged in hybrid human-AI tutoring, using i-Ready and receiving motivational and cognitive support from a remote human tutor. The tutoring intervention took place with tutor-to-student ratios of 1:4. There were 25 active weeks where students engaged in i-Ready only (with an active week defined as 150 or more lesson completions across the population). We removed 13 other inactive weeks that involved sporadic or inconsistent usage resulting from shorter weeks, holidays, state testing interruptions, and other atypical events. Note this analysis of only the active weeks is favorable for the outcomes of these non-tutored weeks such that our results would likely be even stronger if we included them. The next to the last three weeks of school, weeks 40, 41, and 42 (removing the partial week 43 as inactive), students were supported with an early prototype version of a hybrid-human AI tutoring treatment. In this early rough approximation of our vision, remote tutors did not have a dashboard to guide them. Instead, the tutors were advised to visit students through Zoom in a round-robin cycle while students were using i-Ready.

\subsection{Site 3 Method} 
Site 3 took place at a small, urban charter school in Pennsylvania (100\% low income and Black), enrolling only males. None of the 75 students in grades 6-8 were reported to have reached proficiency on state math assessments in the prior academic year (2021-2022), suggesting a low number of prior opportunities for math practice. For the first semester of the school year students engaged in a business-as-usual math curriculum, participating in daily math instruction delivered by their classroom teacher. Weekly MATHia practice is an integrated part of this curriculum. Students began engaging in the treatment of hybrid human-AI tutoring one day per week, in lieu of that day's standard math instruction, at the beginning of the second semester, January 5, 2023. This treatment concluded on May 31, 2023. Hybrid human-AI tutoring was implemented with students engaging in MATHia, while human tutors working remotely provided content-specific and relationship-building support to students. Site 3 was the only site in the current investigation where tutors had access to a real-time tutoring dashboard. MATHia's LiveLab provided tutors real-time information on student engagement and usage by indicating measures of productivity (i.e., workspaces completed) and activity (i.e., idle time).

\section{Results}

\subsection{Site 1 Results}
IXL-reported learning process data of skill usage and proficiency were provided weekly: \textit{time spent}, \textit{questions answered}, and \textit{skills proficient}. Referencing the conceptual pipeline connecting learning process data from human-AI tutoring to changes in student achievement (see Fig. \ref{fig:intro_1}), we classify \textit{time spent} and \textit{questions answered} as step 1 measures and \textit{skills proficient} as a step 2 measure. In addition, students completed Renaissance Learning's Star diagnostic assessment \cite{RenaissanceLearning} up to nine times in the academic year with the first four administrations occurring September, December, February, and March. Descriptive statistics for the September Star diagnostic indicating mean grade level placement and standard deviation are as follows: Control, M = 7.06, SD = 0.133; Delayed Treatment, M = 7.04, SD = 0.070. Both IXL and Star data were collected for math and reading. Table \ref{tab:result_1} displays the site timeline and descriptive statistics of weekly IXL process data for Control and Delayed Treatment groups by time frame (i.e., fall, early spring, and late spring), demonstrating the interrupted time series design \cite{miller2020experimental}.

\begin{table*}[htp]
\centering
\small
\caption{Site 1 timeline of conditions for Control and Delayed Treatment groups with descriptive statistics displaying the math learning process data. The average weekly mean is shown with standard deviation indicated in parenthesis.}
\label{tab:result_1}
\vspace{-0.25cm}
\resizebox{0.85\textwidth}{!}{%
\renewcommand{\arraystretch}{1.2}
\begin{tabular}{llll}
\toprule
& \textbf{Fall} & \textbf{Early Spring} & \textbf{Late Spring} \\
\hline
\multirow{1}{*}{\textbf{Control (n=73)}} & EdTech\_Only & EdTech\_Only & EdTech\_Only \\
\hspace{1em}Time spent (min) & 18.58 (29.72) & 16.20 (23.08) & 6.07 (14.40) \\
\hspace{1em}Questions answered & 43.4 (83.6) & 29.3 (50.5) & 13.9 (36.6) \\
\hspace{1em}Skills proficient & 1.02 (2.31) & 0.688 (1.34) & 0.378 (1.1) \\
\midrule
\multirow{1}{*}{\textbf{Delayed Treatment (n=52)}} & EdTech\_Only &EdTech+MathTeacher&
EdTech+MathTeacher+Tutoring  \\
\hspace{1em}Time spent (min) & 23.48 (36.30) & 37.15 (34.80) & 30.71 (27.83) \\
\hspace{1em}Questions answered & 47.5 (95.6) & 65.3 (80.3) & 47.8 (58.4) \\
\hspace{1em}Skills proficient & 1.08 (2.36) & 1.37 (1.99) & 1.03 (1.29) \\
\bottomrule
\end{tabular}
}

\end{table*}

Independent t-tests were performed comparing IXL process data measures for Control and Delayed Treatment across each time frame (i.e., fall, early spring, late spring) for the following reasons: preliminary data exploration, initial hypothesis testing, and to assist with mixed effects linear regression model development. Preliminary independent t-tests were performed comparing IXL learning process data for Control and Delayed Treatment groups across each time frame (i.e., fall, early spring, late spring). The aim was to determine whether there was a statistically significant difference between each of the learning process measures (dependent variables) across each time frame. Table \ref{tab:result_2} displays the results of the independent t-test. Aligning with our hypothesis that hybrid human-AI tutoring has an effect on student learning, there is a statistically significant difference between the Control and Delayed Treatment groups in early spring and late spring, where the Delayed Treatment students were engaging in EdTech$+$MathTeacher and EdTech$+$MathTeacher\_Tutor conditions, respectively. There were no statistically significant differences between the Control and Delayed Treatment groups in the fall across all learning process measures. Thus, the preliminary independent t-tests confirms statistically significant differences between the use of math software only, EdTech\_Only, compared to with a math teacher, EdTech$+$MathTeacher, and further addition of the hybrid human-AI tutoring, EdTech$+$MathTeacher$+$Tutoring, conditions.

\begin{table*}[h]
\centering
\small
\caption{Independent t-tests comparing Control and Delayed Treatment groups in fall, early spring, and late spring for each average of learning process data per student per week. }
\label{tab:result_2}
\resizebox{0.85\textwidth}{!}{%
\begin{tabular}{lllllll}
\toprule
\textbf{Dependent variable}                  & \textbf{Time frame}   & \textbf{Comparison with Control} & \textbf{t-value} & \textbf{p-value}     & \textbf{CI\_low} & \textbf{CI\_high} \\ \midrule
\multirow{3}{*}{Time spent (min)}         & Fall         & EdTech\_Only                 & 1.71    & 9.02E-02       & -0.9358 & 12.69   \\
                                    & Spring early & EdTech+MathTeacher           & 6.12    & 2.37E-08***  & 13.59  & 26.66    \\
                                    & Spring late  & EdTech+MathTeacher+Tutoring  & 10.7    & 6.89E-17***  & 18.19   & 26.47    \\ \grayline
\multirow{3}{*}{Questions answered} & Fall         & EdTech\_Only                 & 0.70    & 0.480        & -12.23 & 25.67   \\
                                    & Spring early & EdTech+MathTeacher           & 4.72    & 9.26E-06***  & 19.10  & 46.92   \\
                                    & Spring late  & EdTech+MathTeacher+Tutoring  & 6.45    & 5.29E-09***  & 21.09  & 39.86   \\ \grayline
\multirow{3}{*}{Skills proficient}  & Fall         & EdTech\_Only                 & 0.48    & 0.630       & -0.338 & 0.557   \\
                                    & Spring early & EdTech+MathTeacher           & 3.15    & 2.32E-03**   & 0.221  & 0.982   \\
                                    & Spring late  & EdTech+MathTeacher+Tutoring  & 5.53    & 2.18E-07*** & 0.397  & 0.840   \\ \arrayrulecolor{black}\bottomrule 
\multicolumn{7}{l}{{\small *** p<0.001; ** p<0.01; * p<0.05}} \\
\end{tabular}
}
\vspace{-2mm}
\end{table*}

\begin{figure*}[h]
\centering
  \includegraphics[width=0.85\textwidth]{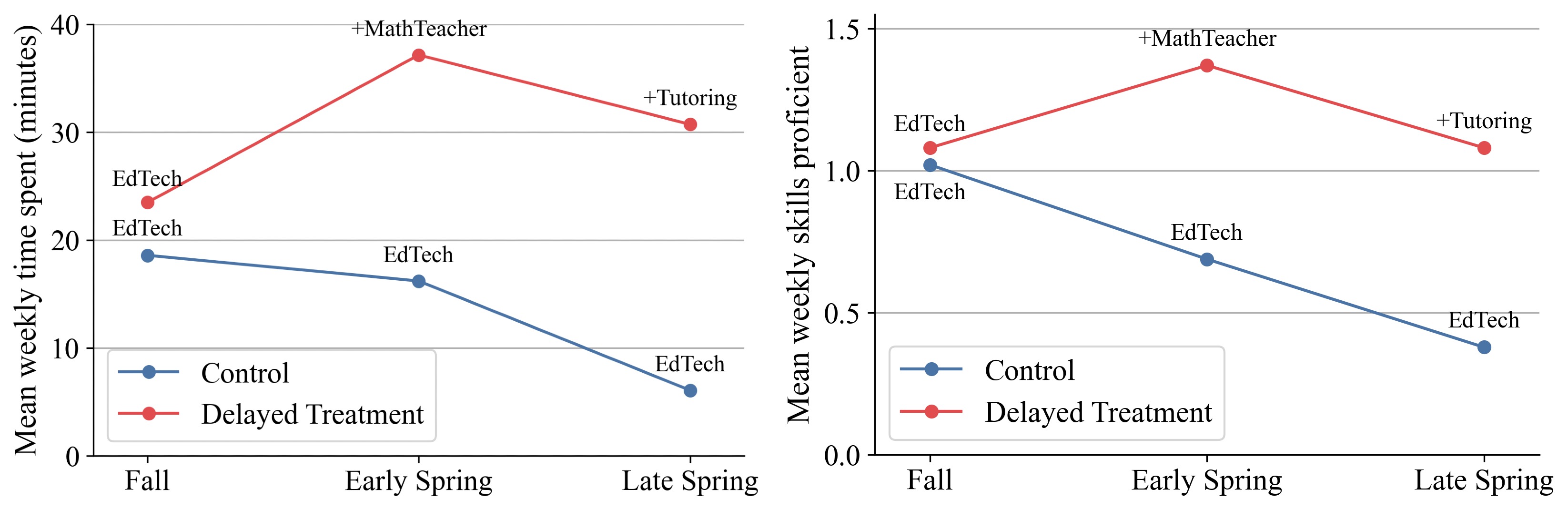}
\caption{Longitudinal plots displaying mean weekly (a) time spent and (b) skills proficient, across fall, early spring, and late spring for Control (blue) and Delayed Treatment (red).}
\label{fig:result_1}
\vspace{-3mm}
\end{figure*}

Longitudinal plots displaying student learning process data for mean weekly time spent and the number of skills for which the student gained proficiency (skills proficient) are shown in Fig. \ref{fig:result_1}a and \ref{fig:result_1}b, respectively. We observe an increase in both measures in early spring for the Delayed Treatment with the addition of a MathTeacher and late spring with Tutoring. To test whether there is a positive and significant effect of Tutoring, with respect to MathTeacher, or if there was a decline, we conducted a mixed effects linear regression.  

A mixed-effects linear model was fitted containing the following fixed-effect predictors: student first Star diagnostic score, taken on or before December 2022 (\textit{pretest}); time period (fall = 00, early spring = 10, late spring = 01); and student group, Control or Delayed Treatment (\textit{student\_group}). The following interaction terms were added: \textit{tutoring:MathTeacher; pretest:tutoring; pretest:MathTeacher}. Individual student (\textit{studentID}) and weeks, numbered 1-40 spanning study duration (\textit{week}), are random effects with fixed means capturing unexplained variability that is unique to each student and week. The generalized mixed model was fit by using Restricted Maximum Likelihood (REML) using R's \texttt{lme4} package (see Equation 1). The model was fitted across all three dependent variables: \textit{time spent, questions answered, and skills proficient}, with all dependent variables and pretest scores normalized.

\vspace{-2mm}
\begin{equation}
\small
\begin{aligned}
Dependent \; variable \sim & pretest + time\_period + student\_group + tutoring * MathTeacher + pretest:tutoring \; + \\
& pretest:MathTeacher + \textit{(1 | \text{studentID})} + \textit{(1 | \text{week})}
\end{aligned}
\end{equation}

As shown in Table \ref{tab:result_3}, we found that hybrid human-AI tutoring had a positive and statistically significant effect on student’s time spent practicing with IXL ($\beta$ = 0.202, 95\% CI:[0.057, 0.347], t(4235) = 2.734, p < .001). Further, the interaction of the pretest and hybrid human-AI tutoring on time spent, \textit{pretest:Tutoring}, demonstrated a negative and statistically significant effect, ($\beta$ = -0.212, 95\% CI:[-0.325, -0.096], t(4235) = 2.734, p < .001). This interaction indicates that the benefits of tutoring were higher for lower pretest students and decreased for higher pretest students.

\begin{table*}[hbp!]
\centering
\small
\caption{Mixed-effects linear model predicting math time spent.}
\vspace{-0.25cm}
\label{tab:result_3}
\resizebox{0.87\textwidth}{!}{%
\renewcommand{\arraystretch}{1.2}
\begin{tabular}{p{3.1cm}P{2.5cm}p{1.5cm}p{1.5cm}p{1.5cm}p{1.5cm}}
\hline
\textbf{Predictors}                 & \textbf{Estimate} & \textbf{SE}     & \textbf{df}    & \textbf{t-value} & \textbf{p-value} \rule{0pt}{2.6ex}\rule[-1.2ex]{0pt}{0pt}      \\ \hline
(intercept)                & 0.017    & 0.0939 & 114   & 0.182   & 0.855748     \\
pretest                    & 0.015    & 0.0465 & 114   & 0.313   & 0.755105     \\
Early\_spring (vs. Fall)    & -0.121   & 0.1114 & 41.05 & -1.089  & 0.282359     \\
Late\_spring (vs. Fall)     & -0.495   & 0.1344 & 41.05 & -3.686  & 0.000660 *** \\
Student\_group\_Control      & -0.048   & 0.0951 & 123.0 & -0.506  & 0.613820     \\
MathTeacher (vs. EdTech)   & 0.487    & 0.0562 & 4334  & 8.670   & < 2e-16 ***  \\
Tutoring (vs. MathTeacher) & 0.202    & 0.0739 & 4325  & 2.734   & 0.006279 **  \\
pretest:MathTeacher        & 0.271    & 0.0444 & 4431  & 6.115   & 1.05e-09 *** \\
pretest:Tutoring           & -0.212   & 0.0592 & 4325  & -3.577  & 0.000351 *** \\ \hline
\multicolumn{5}{l}{{\small *** p<0.001; ** p<0.01; * p<0.05}} 
\end{tabular}
}
\vspace{-0.5cm}
\end{table*}


Perhaps not surprisingly, the addition of a math teacher in the classroom also had a statistically significant positive effect on time spent (see the MathTeacher row in Table \ref{tab:result_3}) and there was a positive interaction with pretest, \textit{pretest:MathTeacher.} This interaction indicates the MathTeacher impact on time was greater for higher pretest students than for lower pretest students, consistent with the left side in Fig. \ref{fig:intro_2}. Oppositely, tutors tend to attend more to low-pretest students (or at least all students uniformly in round robin), indicated by the negative interaction term, \textit{pretest:Tutoring}. A final significant effect of note (see the Late\_spring row in Table \ref{tab:result_3}) is that students’ time spent dropped substantially in the late spring as is also apparent in the Control group lines in Fig. \ref{fig:result_1}.

Effects of the same analysis (see Equation 1) for questions answered and skills proficient as the dependent variable were generally consistent with matching significant results for all but the Tutoring vs. MathTeacher comparison. Referencing RQ1, we found statistically significant positive effects of the MathTeacher for all measures and an additional benefit of human-AI tutoring for time spent. Referencing RQ2, we found that hybrid human-AI tutoring benefited \textit{lower} pretest students more so than \textit{higher} pretest students for all measures, including skills proficient. Conversely, we found that having a math teacher present while students engage in using Edtech benefited \textit{higher} pretest students more so than \textit{lower} pretest students for all measures.  

\vspace{-3mm}

\subsection{Site 2 Results}
In the 25 active weeks prior to engaging in the tutoring treatment, students took an average of 32 minutes to complete a lesson and spent 24 minutes on task each week. During treatment, students took 36 minutes, on average, to complete a lesson and spent 33 minutes on task each week. The increase in student participation due to hybrid human-AI tutoring (33 minutes up from 24 minutes before treatment) approaches the i-Ready data-based recommendation of 45 minutes, even for these historically lower performing students \cite{marple2019iready}. 

\begin{figure*}[h]
\centering
  \includegraphics[width=0.7\textwidth]{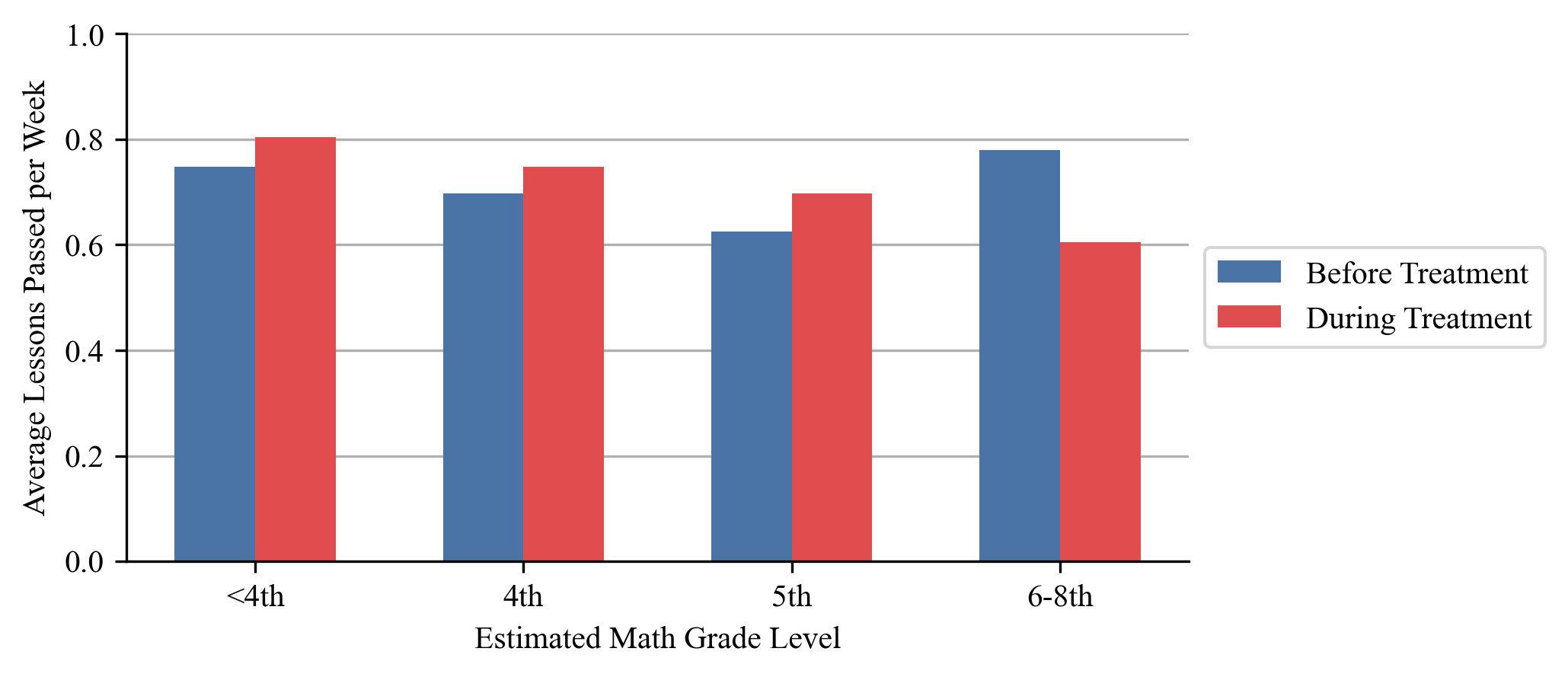}
\vspace{-2mm}
\caption{Average lessons passed per week by estimated grade level. Notice during the hybrid human-AI tutoring treatment, the neediest students (indicated by assigned-lessons being several grades behind), passed more lessons, on average, compared to before treatment.  }
\label{fig:result_2}
\vspace{-3mm}
\end{figure*}

We also analyzed if tutoring support was being distributed to the students with the greatest need. Student's individual need level was operationalized by averaging the grade level of all lessons assigned and completed (both passed and not passed), as determined by the \textit{i-Ready Diagnostic}. Given these diagnostics were computed throughout the school year starting in the fall, the 7th graders at this site are essentially at grade level if they are entering with 6th grade level skills. Fig. \ref{fig:result_2}  displays the average number of lessons passed per week according to students' individual need level, with those below grade level in the left three pairs of bars and those at or above grade level in the rightmost pair. We see that students who were below grade level passed more lessons per week during the last 3 weeks when tutoring was implemented than in the 25 weeks before tutoring whereas the opposite was observed for students at or above grade level. Like Site 1, the Tutoring treatment occurred in late spring and recall there we saw a substantial decline in Control students’ engagement in late spring. We did not have such a control in the Site 2, but if students there also experienced a late spring decline, the positive effects of tutoring were greater than shown in the Fig. \ref{fig:result_2} differences.  

While we cannot adjust for a possible late-spring decline, we can test for the statistical reliability of the differences as shown in Fig. \ref{fig:result_2}. We again followed the general pattern of an interrupted time series analysis \cite{miller2020experimental}. We fitted a mixed effects linear regression model (see Equation 2) with the number of lessons passed per week as the dependent variable and containing the fixed-effect predictors: average grade level across all lessons completed by a student (\textit{lesson\_grade}); and whether the week was before or during treatment (\textit{during\_treatment}). Individual student (\textit{studentID}) is a random effect, accounting for the correlation between lessons passed by a student across weeks. The \textit{lesson\_grade} serves as a proxy of students' prior math knowledge expressed in grade level (e.g., a student with an average of 4.5 is estimated to have math knowledge comparable to a student halfway through 4th grade). Parameter estimates are shown in Table \ref{tab:result_5}.

\begin{equation}
Lessons \; passed \; per \; week \sim lesson\_grade * during\_treatment + \textit{(1 | \text{studentID})}
\end{equation}

\begin{table*}[h]
\centering
\small
\caption{Mixed-effects linear model predicting lessons passed per week.}
\label{tab:result_5}
\resizebox{0.95\textwidth}{!}{%
\renewcommand{\arraystretch}{1.2}
\begin{tabular}{p{4cm}P{2.5cm}p{1.5cm}p{1.5cm}p{1.5cm}p{1.5cm}}
\hline
\textbf{Predictors}                 & \textbf{Estimate} & \textbf{SE}     & \textbf{df}    & \textbf{t-value} & \textbf{p-value} \rule{0pt}{2.6ex}\rule[-1.2ex]{0pt}{0pt}      \\ \hline
(intercept)                        & 6.880e-01  & 5.527e-02 & 3.955e+02 & 12.449 & \textless 2e-16 *** \\
lesson\_grade                      & 6.004e-03  & 1.119e-02 & 3.955e+02 & 0.537  & 0.591908            \\
during\_treatment\_Y               & 2.437e-01  & 7.022e-02 & 1.034e+04 & 3.470  & 0.000522 ***        \\
lesson\_grade:during\_treatment\_Y & -5.261e-02 & 1.422e-02 & 1.034e+04 & -3.700 & 0.000217 ***       \\ \hline
\multicolumn{5}{l}{{\small \textit{Note:} *** p<0.001; ** p<0.01; * p<0.05}} 
\end{tabular}
}
\vspace{-0.25cm}
\end{table*}

There was a positive and statistically significant main effect for the number of lessons students passed per week during the treatment, $\beta$ = 2.44e-01, CI:[0.106 , 0.381], t(10300) = 3.47, p < 0.001. Students completed, on average, more lessons each week during the three weeks of the human-AI tutoring treatment. There was also a negative and statistically significant interaction between student’s estimated grade level and tutoring treatment, demonstrating the student’s most in need are passing more lessons while participating in tutoring treatment. While students in general benefited from tutoring (and effects may be even bigger if there is a general late spring decline), students at or above grade level did not benefit. The round-robin tutoring approach we used, given dashboard guidance was not yet available, may have led to unneeded interruptions that distracted already-engaged students from making progress in the software.

\subsection{Site 3 Results}

\begin{figure*}[h]
\centering
  \includegraphics[width=0.45\textwidth]{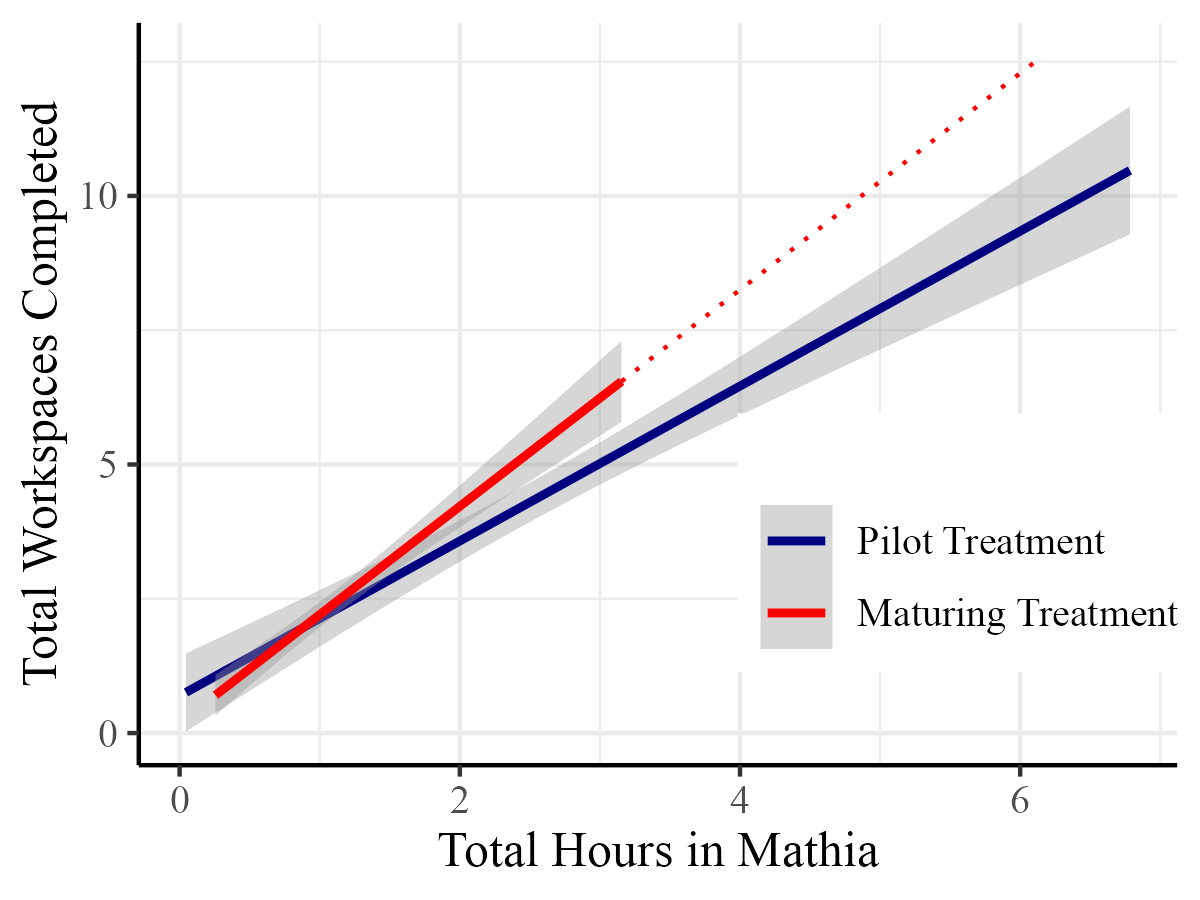}
\caption{Total workspaces (i.e., lessons) completed compared to the total time each student spent using MATHia while engaging in the treatment involving tutor-to-student ratios during the Pilot Treatment phase (1:8, blue) and Maturing Treatment phase (1:4, red). The projected student progress while engaging in the treatment with tutor-to-student ratios of 1:4 is indicated by the dotted red line.}
\label{fig:result_3}
\vspace{-6mm}
\end{figure*}

Our early attempts at remote tutoring started at Site 3 and, as a consequence of our early-stage efforts, the implementation evolved over the course of the Spring semester. Our analysis thus explores the differences in impact of  the early phase of implementation with that of the later phase. The intervention changed most substantially in two ways. First, time spent tutoring students increased from about 25 to 50 minutes per session; however, this also coincided with a shift in frequency of tutoring from once a week to once every other week such that available tutoring time over two weeks was the same. Second, the number of tutors available per class increased with tutor to student ratios changing from about 1:8 to about 1:4. Given this shift, we considered how increases in tutor availability and longer sessions may have affected student progress in MATHia. We measured student progress as the average number of MATHia workspace lessons completed per hour. Fig. \ref{fig:result_3} illustrates the total workspace lessons completed compared to the total time using MATHia for students while engaging in the treatment, involving tutor-to-student ratios of 1:8 (blue) and 1:4 (red). The rate of progress increases for students upon shifting from tutor-to-student ratios of 1:8 to 1:4. Students completed an additional 0.36 (95\% CI:[0.02, 0.70]) workspaces per hour of MATHia engagement \textit{after} the shift when they had greater access to hybrid human-AI tutoring, t(131.15) = 2.24, p = .03, $\beta_{0}$ = 1.96.\footnote{One student was excluded from analysis for lack of usage (less than 3 minutes) before the shift in tutoring.}

\section{Discussion}

\textbf{Human-AI tutoring has positive impacts on learning process data and student outcomes.} Across all three sites, our findings support the hypothesis that hybrid human-AI tutoring increases student engagement with AI software and learning progresses compared to student AI-software use alone (RQ1). In Site 1, we find statistically significant increases in average \textit{time spent} and \textit{skills proficient} among students participating in EdTech$+$MathTeacher and EdTech$+$MathTeacher$+$Tutoring compared to the Control (EdTech\_Only) group (see Table \ref{tab:result_2}). Mixed-effects linear models confirm that the EdTech$+$MathTeacher treatment increases engagement as measured by time spent, and that the EdTech$+$MathTeacher$+$Tutoring treatment provides a statistically significant (though smaller) additional boost (see Table \ref{tab:result_3}). The EdTech$+$MathTeacher treatment also increases the number of lessons completed in the math software, relative to the control group, but the additional boost provided by the EdTech$+$MathTeacher$+$Tutoring treatment is not statistically significant.  In Site 2, we find human-AI tutoring increased engagement with AI in terms of time spent from 24 to 33 minutes per week and also boosted lesson completion rates for students whose skills assessments suggested greater need. Interestingly, human-AI tutoring may have actually slowed the progress of students who were at or above grade level (See Fig. \ref{fig:result_2}). This finding hints at the possibility that  the ``round-robin'' approach that enforced equal participation in human tutoring among all students could be productively replaced by a system that directs tutor effort toward students with greater need (e.g., where the tutor decides whom to help, based on information on a dashboard). In Site 3, we took a step in this direction by piloting the introduction of a live tutoring dashboard into a human-AI tutoring intervention. Variations of the human-AI intervention were provided to all students, so this site lacks a Control group, \textit{per se}.  We found students completed more work spaces per hour of participation (compared to working with the math learning software without a human tutor) in the intervention with tutor-student ratios of 1:4 rather than 1:8.

Demonstrating benefits of hybrid tutoring using intermediate measures like time spent or skills proficient within the math software is a valuable step. Nevertheless, we also want to know how hybrid AI-human tutoring impacts students’ performance on external assessments of knowledge, such as state standardized tests.  Referring to the conceptual pathway in Fig. \ref{fig:intro_1}, linking human-AI tutoring to higher achievement, our current results provide strong support for step 1 (increased learning opportunities) and somewhat more tentative support for step 2 (increases in math software lesson completion). We do not present new results on step 3 (increases in achievement). Instead, we draw upon results from prior studies linking increases in learning opportunities and lesson completion achieved through recommended use of our software tools to realized increases in state test scores associated with that recommended level of use. We then use these past correlations to predict the achievement gains that could result from the increased utilization of software tools induced by our human-AI tutoring intervention. Table \ref{tab:disc_1} displays an overview of the software applications used in this study, the documented past correlations between ``recommended'' use of this software and subsequent test score gains, and the increased utilization of the software associated with the human-AI tutoring interventions demonstrated in this study.

For Site 1 using IXL, the average weekly skills proficient for both groups participating in math software use only was 0.88 (see 1st row, 4th column in Table \ref{tab:disc_1}), whereas students engaging in hybrid human-AI tutoring as treatment averaged 1.03 skills proficient per week (5th column). To put this finding into perspective, a prior IXL-conducted study indicates students who reach proficiency on one IXL math skill \textit{every other week} demonstrate statistically significant increases in performance on the Pennsylvania State Assessment (PSSA) after a two- year period \cite{ixleffect2023}. At our study site, treatment students attain, on average, at least one skill proficient every week. Evidence from Site 1, when considered alongside prior IXL research \cite{ixleffect2023}, tentatively suggests that hybrid tutoring may contribute to enhanced student performance, indicating the potential of accelerated achievement gains.

\begin{table*}[htp!]
\centering
\small
\caption{Overview of software used with the research-supported recommendation correlating with predicted outcomes on state standardized tests and our findings for student software use only and hybrid human-AI tutoring.}
\label{tab:disc_1}

\renewcommand{\arraystretch}{1.3}
\begin{tabular}{m{0.1\columnwidth}p{0.23\columnwidth}p{0.23\columnwidth}p{0.13\columnwidth}p{0.17\columnwidth}}
\toprule
\textbf{Software} & \textbf{Recommended level of use} & \textbf{Predicted outcome} & \textbf{Software only} & \textbf{Human-AI tutoring (treatment)} \\ 
\midrule
\multirow{1}{*}{IXL} & 1 skill proficient every other week & Significantly better on PSSA, 9-point percentile gain after 2 years of practice* & 0.88 skills proficient/wk & 1.03 skills proficient/wk \\ 
\grayline
i-Ready & 45 min./week & Significant improvement on SBAC** & 24 min./wk & 33 min./wk \\ 
\grayline
\multirow{1}{*}{MATHia} & 60 min./week + 0.5-0.75 mastery workspaces passed/hour & Significant improvement on various standardized test(s)*** & N/A & 0.36 more workspaces passed/hr  \\ \arrayrulecolor{black}\bottomrule 
\end{tabular}


\hspace{-30.5em}\small\footnotesize{*Pennsylvania System of School Assessment, two-year study, \cite{ixleffect2023}}\\
\hspace{-33.6em}\small\footnotesize{**Smarter Balanced Assessments Consortium Math Assessment, \cite{marple2019iready}}\\
\footnotesize ***CTB/McGraw Hill Acuity Series \cite{sales2015exploring, pane2014effectiveness}; additional correlational studies associate MATHia usage with state test outcomes in Virginia and Florida \cite{zheng2019using}.

\end{table*}

At Site 2, which used i-Ready, all students engaged in i-Ready for the majority of the academic year, averaging 24 minutes per week (see 2nd row, 4th column in Table \ref{tab:disc_1}). In the later three weeks of hybrid human-AI tutoring treatment, average usage increased to 33 minutes per week. Curriculum Associates reports 45 minutes of usage per week is associated with significant improvements on the SBA \cite{marple2019iready} and suggests 30-49 minutes of time-on-task weekly with at least 70 percent of lessons passed for the year \cite{CurriculumAssociatesFeb2023}. Our results suggest that human-AI tutoring generates substantial progress towards that recommended level of utilization.

Site 3 used MATHia but always combined with remote human tutoring, thus we cannot compare human-AI tutoring to software only. However, we did see an impact from our improved remote tutoring implementation, which particularly changed in going from to a tutor-student ratio of 1:8 to 1:4. This change led to students passing 0.36 more workspace lessons per hour than they did earlier (3rd row, 5th column in Table \ref{tab:disc_1}). This increase in workplace pass rate (above the unknown software only rate) likely puts students well within MATHia’s recommended 0.5-0.75 workspaces passed per hour recommendation (1st column).  MATHia also recommends 60 minutes of use per week. We hope to encourage Site 3 to use human-AI tutoring every week instead of other week to better approach this recommendation.

\textbf{Students with the greatest needs benefit more from hybrid human-AI tutoring than students meeting grade level.} Although we find hybrid human-AI tutoring to have positive effects on students’ engagement in learning, these effects are not uniform across students of varying abilities (RQ2). Our mixed effects linear model analysis at Site 1 identified an interaction between pretest and the impact of adding human tutoring support in conjunction with EdTech use. For all measures of time spent, questions answered, and skills proficient, human-AI tutoring produced significantly better outcomes for lower pre-test students than for higher pre-test students. We found a similar result at Site 2. For students below grade level (as identified by i-Ready’s diagnostic score), the addition of human-AI tutoring raised their lesson completion relative to EdTech use alone, however, for students at or above grade level, the addition of human-AI tutoring may have slowed their lesson completion. Note that because we could not estimate a late-spring decline in Site 2, it may be that human-AI tutoring also helped students at or above grade level. Even if the human-AI tutoring effect at Site 2 is generally higher than estimated, the interaction effect remains. Among a majority of low-income students, our analysis indicates that lower-performing students, identified by lower pretest scores, may gain more than higher-performing students. Human-AI tutoring is benefiting the students most in need of it. Why might this be?

This result is just what we expect from dashboard-driven tutoring (right side in Fig \ref{fig:intro_2}), but we were surprised to see it as a consequence of the round-robin tutoring approach used at Sites 1 and 2. While tutors did not visit needier, lower pre-test students more often, they may have stayed longer with them. A second possible explanation is that student awareness that a tutor is coming has a positive impact on engagement particularly for students who have a history of lower engagement. A third is that students with lower pretests are likely to need and benefit more from the relationship building and human tutor interaction provided. These results hint at the possibility that a dashboard-directed system that pushes tutors away from a ``round-robin'' engagement with all students to even greater interaction with students below grade level could yield even stronger aggregate results (cf. \cite{holstein2018student}) and achieve greater educational equity.

\textbf{Math teacher support during EdTech use appears beneficial but may not be ideally equitable.} Referencing the findings for RQ2, math teachers play a significant role in student engagement and learning when working in conjunction with AI-driven math software (cf. \cite{holstein2018student}). The importance of their role is evidenced in Site 1, where the EdTech$+$MathTeacher treatment yields statistically significant positive effects on math time spent and skills proficient relative to the Control condition without a math teacher. Perhaps the supervisor available in the Control group throughout and available in the Delayed Treatment in the fall did not provide any extra tutoring support beyond the software whereas the Math Teacher did. Our analyses further indicated that having a math teacher present while students engage in using Edtech benefited higher pretest more than lower pretest students. This result may be a consequence of teachers being inclined to support students who request help and higher prior achieving students are more likely to do so. Better understanding of these teacher effects on student engagement and educational equity  is an interesting direction for future work.

\textbf{Rapid, quasi-experiments provide early evaluation guidance.} We suggest quasi-experimental methods toward collecting rapid and reasonably reliable evidence both to guide iterative design in early program development and to create a robust ``web of validity'' as a program matures. These methods can not only address intervention effectiveness but also delve into the mechanisms behind its effect. Our results support the notion that human-AI tutoring can increase engagement with math software and some intermediate measures of student learning, that this effect exists especially for students below grade level, and that low tutor to student ratios dampen impact. This approach provides useful data early in a project without the high costs, in time and money, required by traditional RCTs.

\section{Limitations, Implications, Future Work, \& Conclusion}


While these quasi-experiments shed light on the effect of hybrid human-AI tutoring during the school day, they also possess  important limitations. The absence of random assignment introduces the possibility of selection bias and limited control over extraneous variables not present in true experiments \cite{miller2020experimental}. Small sample sizes also reduce statistical power, hindering analysis of more complicated interactions. Implementation fidelity is harder to determine in these regular math classroom-based field studies than in an after-school program settings over which researchers might be able to exercise more control. Each site differed in implementation details (see Table \ref{tab:met_1}) which may limit the external validity of our results, as school-specific factors and demographics may influence human-AI tutoring intervention’s effectiveness differently in other settings. Finally, we employed disparate measures across sites, which may raise issues concerning validity and reliability. The measures used reflect the constraints and recommendations (see Table 7) of the different ed-tech tools. Where possible, we used consistent measures (e.g., time). Using different math software enhances the generalizability of the findings reflecting the reality of implementing interventions aimed at helping students who need it the most. Remaining software-agnostic enhances equity by increasing access to our intervention among students and schools. Despite these limitations, our findings provide valuable insights on how to develop hybrid human-AI tutoring models, which will guide subsequent efforts to confirm and extend our findings. 

This work suggests many productive areas for future research, including developing refined models of hybrid human-AI tutoring to direct attention to those who can benefit most (such as through AI dashboard support), and conducting control trials to understand the impact of hybrid human-AI tutoring on external assessments. Another key question for broader impact is to consider further refinements to increase the cost-effectiveness of this model. Future work will also investigate the heterogeneity of treatment effects across student-level demographics.

Averaging across our three sites our marginal cost per student was below \$750, with a range from \$597 to  \$1170 per year.\footnote{Our costs reflect tutor compensation set at \$15/hr, which is more than twice the state minimum wage in Pennsylvania.  Programs that rely on volunteer tutors or Americorps members could attain significantly lower cost levels.} The variability in annual cost is related to software licensing fees and site-based administration costs. Our work to date thus suggests that a marginal cost of about \$700 per student, which is a small fraction of the recorded \$3,500-\$4,300 per student costs associated with other high-impact tutoring programs \cite{guryan2021not}, is attainable. The powerful combination of potentially high efficacy and low marginal costs suggested by our early-stage results holds out the exciting possibility that this line of research could eventually yield high social returns.

\begin{acks}
The authors thank Cindy Tipper and Hui Cheng for their dedicated and considerable contribution in data retrieval and analysis. They also extend their appreciation to Curriculum Associates, LLC, for their partnership and collaborative efforts in this research. This work was made possible with support from the Richard King Mellon Foundation (Grant No. 12126) and the Learning Engineering Virtual Institute. Any opinions, findings, and conclusions expressed in this material are those of the authors.
\end{acks}

\bibliographystyle{ACM-Reference-Format}
\bibliography{source/lak24-53}


\begin{thebibliography}{33}


\ifx \showCODEN    \undefined \def \showCODEN     #1{\unskip}     \fi
\ifx \showDOI      \undefined \def \showDOI       #1{#1}\fi
\ifx \showISBNx    \undefined \def \showISBNx     #1{\unskip}     \fi
\ifx \showISBNxiii \undefined \def \showISBNxiii  #1{\unskip}     \fi
\ifx \showISSN     \undefined \def \showISSN      #1{\unskip}     \fi
\ifx \showLCCN     \undefined \def \showLCCN      #1{\unskip}     \fi
\ifx \shownote     \undefined \def \shownote      #1{#1}          \fi
\ifx \showarticletitle \undefined \def \showarticletitle #1{#1}   \fi
\ifx \showURL      \undefined \def \showURL       {\relax}        \fi
\providecommand\bibfield[2]{#2}
\providecommand\bibinfo[2]{#2}
\providecommand\natexlab[1]{#1}
\providecommand\showeprint[2][]{arXiv:#2}

\bibitem[Accelerator(2021)]%
        {accelerator2021toolkit}
\bibfield{author}{\bibinfo{person}{National Student~Support Accelerator}.} \bibinfo{year}{2021}\natexlab{}.
\newblock \bibinfo{title}{Toolkit for Tutoring Programs}.
\newblock
\newblock
\urldef\tempurl%
\url{https://doi.org/10.26300/5n7h-mh59}
\showDOI{\tempurl}


\bibitem[Aleven et~al\mbox{.}(2023)]%
        {aleven2023towards}
\bibfield{author}{\bibinfo{person}{Vincent Aleven}, \bibinfo{person}{Richard Baraniuk}, \bibinfo{person}{Emma Brunskill}, \bibinfo{person}{Scott Crossley}, \bibinfo{person}{Dora Demszky}, \bibinfo{person}{Stephen Fancsali}, \bibinfo{person}{Shivang Gupta}, \bibinfo{person}{Kenneth Koedinger}, \bibinfo{person}{Chris Piech}, \bibinfo{person}{Steve Ritter}, {et~al\mbox{.}}} \bibinfo{year}{2023}\natexlab{}.
\newblock \showarticletitle{Towards the Future of AI-Augmented Human Tutoring in Math Learning}. In \bibinfo{booktitle}{\emph{International Conference on Artificial Intelligence in Education}}. \bibinfo{publisher}{Springer}, \bibinfo{address}{Tokyo, Japan}, \bibinfo{pages}{26--31}.
\newblock


\bibitem[Aleven et~al\mbox{.}(2017)]%
        {aleven2016instruction}
\bibfield{author}{\bibinfo{person}{Vincent Aleven}, \bibinfo{person}{Elizabeth~A McLaughlin}, \bibinfo{person}{R~Amos Glenn}, {and} \bibinfo{person}{Kenneth~R Koedinger}.} \bibinfo{year}{2017}\natexlab{}.
\newblock \showarticletitle{Instruction based on adaptive learning technologies}.
\newblock \bibinfo{journal}{\emph{Handbook of research on learning and instruction}}  \bibinfo{volume}{2} (\bibinfo{year}{2017}), \bibinfo{pages}{522--560}.
\newblock


\bibitem[Alliance(2020)]%
        {afterschool2020}
\bibfield{author}{\bibinfo{person}{Afterschool Alliance}.} \bibinfo{year}{2020}\natexlab{}.
\newblock \bibinfo{title}{America After 3PM: Demand Grows, Opportunity Shrinks}.
\newblock
\newblock


\bibitem[Associates(2023)]%
        {CurriculumAssociatesFeb2023}
\bibfield{author}{\bibinfo{person}{Curriculum Associates}.} \bibinfo{year}{February, 2023}\natexlab{}.
\newblock \bibinfo{booktitle}{\emph{Impact of i-Ready Personalized Instruction with Fidelity on Student Achievement in Mathematics}}.
\newblock \bibinfo{type}{{T}echnical {R}eport}.
\newblock
\urldef\tempurl%
\url{https://www.curriculumassociates.com/-/media/mainsite/files/i-ready/ca-mcas-impact-of-iready-math-2023.pdf}
\showURL{%
\tempurl}


\bibitem[Chhabra et~al\mbox{.}(2022)]%
        {chhabra2022evaluation}
\bibfield{author}{\bibinfo{person}{Pallavi Chhabra}, \bibinfo{person}{Danielle Chine}, \bibinfo{person}{Adetunji Adeniran}, \bibinfo{person}{Shivang Gupta}, {and} \bibinfo{person}{Kenneth Koedinger}.} \bibinfo{year}{2022}\natexlab{}.
\newblock \showarticletitle{An Evaluation of Perceptions Regarding Mentor Competencies for Technology-based Personalized Learning}. In \bibinfo{booktitle}{\emph{Society for Information Technology \& Teacher Education International Conference}}. \bibinfo{publisher}{Association for the Advancement of Computing in Education (AACE)}, \bibinfo{address}{Waynesville, NC USA}, \bibinfo{pages}{1812--1817}.
\newblock


\bibitem[Chine et~al\mbox{.}(2022)]%
        {chine2022educational}
\bibfield{author}{\bibinfo{person}{Danielle~R Chine}, \bibinfo{person}{Cassandra Brentley}, \bibinfo{person}{Carmen Thomas-Browne}, \bibinfo{person}{J~Elizabeth Richey}, \bibinfo{person}{Abdulmenaf Gul}, \bibinfo{person}{Paulo~F Carvalho}, \bibinfo{person}{Lee Branstetter}, {and} \bibinfo{person}{Kenneth~R Koedinger}.} \bibinfo{year}{2022}\natexlab{}.
\newblock \showarticletitle{Educational equity through combined human-AI personalization: A propensity matching evaluation}. In \bibinfo{booktitle}{\emph{International Conference on Artificial Intelligence in Education}}. \bibinfo{publisher}{Springer}, \bibinfo{address}{Durham, UK}, \bibinfo{pages}{366--377}.
\newblock


\bibitem[Corbett et~al\mbox{.}(2001)]%
        {corbett2001cognitive}
\bibfield{author}{\bibinfo{person}{Albert~T Corbett}, \bibinfo{person}{Kenneth Koedinger}, {and} \bibinfo{person}{William~S Hadley}.} \bibinfo{year}{2001}\natexlab{}.
\newblock \showarticletitle{Cognitive tutors: From the research classroom to all classrooms}.
\newblock In \bibinfo{booktitle}{\emph{Technology enhanced learning}}. \bibinfo{publisher}{Routledge}, \bibinfo{pages}{215--240}.
\newblock


\bibitem[Curriculumassociates.com(2023)]%
        {curriculumassociates2023}
\bibfield{author}{\bibinfo{person}{Curriculumassociates.com}.} \bibinfo{year}{2023}\natexlab{}.
\newblock \bibinfo{title}{Curriculum Associates}.
\newblock
\newblock
\urldef\tempurl%
\url{https://www.curriculumassociates.com/}
\showURL{%
\tempurl}


\bibitem[Deaton and Cartwright(2018)]%
        {deaton2018understanding}
\bibfield{author}{\bibinfo{person}{Angus Deaton} {and} \bibinfo{person}{Nancy Cartwright}.} \bibinfo{year}{2018}\natexlab{}.
\newblock \showarticletitle{Understanding and misunderstanding randomized controlled trials}.
\newblock \bibinfo{journal}{\emph{Social science \& medicine}}  \bibinfo{volume}{210} (\bibinfo{year}{2018}), \bibinfo{pages}{2--21}.
\newblock


\bibitem[Escueta et~al\mbox{.}(2020)]%
        {escueta2020upgrading}
\bibfield{author}{\bibinfo{person}{Maya Escueta}, \bibinfo{person}{Andre~Joshua Nickow}, \bibinfo{person}{Philip Oreopoulos}, {and} \bibinfo{person}{Vincent Quan}.} \bibinfo{year}{2020}\natexlab{}.
\newblock \showarticletitle{Upgrading education with technology: Insights from experimental research}.
\newblock \bibinfo{journal}{\emph{Journal of Economic Literature}} \bibinfo{volume}{58}, \bibinfo{number}{4} (\bibinfo{year}{2020}), \bibinfo{pages}{897--996}.
\newblock


\bibitem[Guryan et~al\mbox{.}(2021)]%
        {guryan2021not}
\bibfield{author}{\bibinfo{person}{Jonathan Guryan}, \bibinfo{person}{Jens Ludwig}, \bibinfo{person}{Monica~P Bhatt}, \bibinfo{person}{Philip~J Cook}, \bibinfo{person}{Jonathan Davis}, \bibinfo{person}{Kenneth Dodge}, \bibinfo{person}{George Farkas}, \bibinfo{person}{Roland~G Fryer~Jr}, \bibinfo{person}{Susan Mayer}, \bibinfo{person}{Harold Pollack}, {et~al\mbox{.}}} \bibinfo{year}{2021}\natexlab{}.
\newblock \showarticletitle{Not Too Late: Improving Academic Outcomes among Adolescents (Working Paper 28531).}
\newblock \bibinfo{journal}{\emph{National Bureau of Economic Research}} (\bibinfo{year}{2021}).
\newblock


\bibitem[Hariton and Locascio(2018)]%
        {hariton2018randomised}
\bibfield{author}{\bibinfo{person}{Eduardo Hariton} {and} \bibinfo{person}{Joseph~J Locascio}.} \bibinfo{year}{2018}\natexlab{}.
\newblock \showarticletitle{Randomised controlled trials—the gold standard for effectiveness research}.
\newblock \bibinfo{journal}{\emph{BJOG: an international journal of obstetrics and gynaecology}} \bibinfo{volume}{125}, \bibinfo{number}{13} (\bibinfo{year}{2018}), \bibinfo{pages}{1716}.
\newblock


\bibitem[Heffernan and Heffernan(2014)]%
        {heffernan2014assistments}
\bibfield{author}{\bibinfo{person}{Neil~T Heffernan} {and} \bibinfo{person}{Cristina~Lindquist Heffernan}.} \bibinfo{year}{2014}\natexlab{}.
\newblock \showarticletitle{The ASSISTments ecosystem: Building a platform that brings scientists and teachers together for minimally invasive research on human learning and teaching}.
\newblock \bibinfo{journal}{\emph{International Journal of Artificial Intelligence in Education}}  \bibinfo{volume}{24} (\bibinfo{year}{2014}), \bibinfo{pages}{470--497}.
\newblock


\bibitem[Holstein and Aleven(2022)]%
        {holstein2022designing}
\bibfield{author}{\bibinfo{person}{Kenneth Holstein} {and} \bibinfo{person}{Vincent Aleven}.} \bibinfo{year}{2022}\natexlab{}.
\newblock \showarticletitle{Designing for human--AI complementarity in K-12 education}.
\newblock \bibinfo{journal}{\emph{AI Magazine}} \bibinfo{volume}{43}, \bibinfo{number}{2} (\bibinfo{year}{2022}), \bibinfo{pages}{239--248}.
\newblock


\bibitem[Holstein et~al\mbox{.}(2017)]%
        {holstein2017intelligent}
\bibfield{author}{\bibinfo{person}{Kenneth Holstein}, \bibinfo{person}{Bruce~M McLaren}, {and} \bibinfo{person}{Vincent Aleven}.} \bibinfo{year}{2017}\natexlab{}.
\newblock \showarticletitle{Intelligent tutors as teachers' aides: exploring teacher needs for real-time analytics in blended classrooms}. In \bibinfo{booktitle}{\emph{Proceedings of the seventh international learning analytics \& knowledge conference}}. \bibinfo{publisher}{ACM}, \bibinfo{address}{Vancouver, Canada}, \bibinfo{pages}{257--266}.
\newblock


\bibitem[Holstein et~al\mbox{.}(2018)]%
        {holstein2018student}
\bibfield{author}{\bibinfo{person}{Kenneth Holstein}, \bibinfo{person}{Bruce~M McLaren}, {and} \bibinfo{person}{Vincent Aleven}.} \bibinfo{year}{2018}\natexlab{}.
\newblock \showarticletitle{Student learning benefits of a mixed-reality teacher awareness tool in AI-enhanced classrooms}. In \bibinfo{booktitle}{\emph{Artificial Intelligence in Education: 19th International Conference, AIED 2018, London, UK, June 27--30, 2018, Proceedings, Part I 19}}. \bibinfo{publisher}{Springer}, \bibinfo{address}{London, UK}, \bibinfo{pages}{154--168}.
\newblock


\bibitem[Koedinger et~al\mbox{.}(2023)]%
        {koedinger2023astonishing}
\bibfield{author}{\bibinfo{person}{Kenneth~R Koedinger}, \bibinfo{person}{Paulo~F Carvalho}, \bibinfo{person}{Ran Liu}, {and} \bibinfo{person}{Elizabeth~A McLaughlin}.} \bibinfo{year}{2023}\natexlab{}.
\newblock \showarticletitle{An astonishing regularity in student learning rate}.
\newblock \bibinfo{journal}{\emph{Proceedings of the National Academy of Sciences}} \bibinfo{volume}{120}, \bibinfo{number}{13} (\bibinfo{year}{2023}), \bibinfo{pages}{e2221311120}.
\newblock


\bibitem[Kraft et~al\mbox{.}(2021)]%
        {kraft2021school}
\bibfield{author}{\bibinfo{person}{Matthew~A Kraft}, \bibinfo{person}{Alexander Bolves}, {and} \bibinfo{person}{Noelle~M Hurd}.} \bibinfo{year}{2021}\natexlab{}.
\newblock \showarticletitle{School-based mentoring relationships and human capital formation}.
\newblock \bibinfo{journal}{\emph{EdWorkingPaper: 21}}  \bibinfo{volume}{441} (\bibinfo{year}{2021}).
\newblock


\bibitem[Kraft and Falken(2021)]%
        {kraft2021blueprint}
\bibfield{author}{\bibinfo{person}{Matthew~A Kraft} {and} \bibinfo{person}{Grace~T Falken}.} \bibinfo{year}{2021}\natexlab{}.
\newblock \bibinfo{title}{A blueprint for scaling tutoring across public schools (EdWorkingPaper No. 21--335). Annenberg Institute at Brown University}.
\newblock
\newblock


\bibitem[Learning(2023)]%
        {RenaissanceLearning}
\bibfield{author}{\bibinfo{person}{Renaissance Learning}.} \bibinfo{year}{2023}\natexlab{}.
\newblock
\newblock
\urldef\tempurl%
\url{https://www.renaissance.com/}
\showURL{%
\tempurl}


\bibitem[Marple et~al\mbox{.}(2019)]%
        {marple2019iready}
\bibfield{author}{\bibinfo{person}{S. Marple}, \bibinfo{person}{K. Jaquet}, \bibinfo{person}{A. Laudone}, \bibinfo{person}{J. Sewell}, {and} \bibinfo{person}{K. Liepmann}.} \bibinfo{year}{2019}\natexlab{}.
\newblock \bibinfo{booktitle}{\emph{i-Ready in 7th Grade Math Classes: A Mixed Methods Case Study}}.
\newblock \bibinfo{type}{{T}echnical {R}eport}. \bibinfo{institution}{WestEd}, \bibinfo{address}{San Francisco, CA}.
\newblock
\urldef\tempurl%
\url{https://www.curriculumassociates.com/-/media/mainsite/files/i-ready/i-ready-study-full-report_final.pdf}
\showURL{%
\tempurl}


\bibitem[Miller et~al\mbox{.}(2020)]%
        {miller2020experimental}
\bibfield{author}{\bibinfo{person}{Christopher~J Miller}, \bibinfo{person}{Shawna~N Smith}, {and} \bibinfo{person}{Marianne Pugatch}.} \bibinfo{year}{2020}\natexlab{}.
\newblock \showarticletitle{Experimental and quasi-experimental designs in implementation research}.
\newblock \bibinfo{journal}{\emph{Psychiatry research}}  \bibinfo{volume}{283} (\bibinfo{year}{2020}), \bibinfo{pages}{112452}.
\newblock


\bibitem[Muralidharan et~al\mbox{.}(2019)]%
        {muralidharan2019disrupting}
\bibfield{author}{\bibinfo{person}{Karthik Muralidharan}, \bibinfo{person}{Abhijeet Singh}, {and} \bibinfo{person}{Alejandro~J Ganimian}.} \bibinfo{year}{2019}\natexlab{}.
\newblock \showarticletitle{Disrupting education? Experimental evidence on technology-aided instruction in India}.
\newblock \bibinfo{journal}{\emph{American Economic Review}} \bibinfo{volume}{109}, \bibinfo{number}{4} (\bibinfo{year}{2019}), \bibinfo{pages}{1426--1460}.
\newblock


\bibitem[Nickow et~al\mbox{.}(2020)]%
        {nickow2020impressive}
\bibfield{author}{\bibinfo{person}{Andre Nickow}, \bibinfo{person}{Philip Oreopoulos}, {and} \bibinfo{person}{Vincent Quan}.} \bibinfo{year}{2020}\natexlab{}.
\newblock \showarticletitle{The impressive effects of tutoring on prek-12 learning: A systematic review and meta-analysis of the experimental evidence}.
\newblock \bibinfo{journal}{\emph{NBER Working Papers}} (\bibinfo{year}{2020}).
\newblock


\bibitem[Pane et~al\mbox{.}(2014)]%
        {pane2014effectiveness}
\bibfield{author}{\bibinfo{person}{John~F Pane}, \bibinfo{person}{Beth~Ann Griffin}, \bibinfo{person}{Daniel~F McCaffrey}, {and} \bibinfo{person}{Rita Karam}.} \bibinfo{year}{2014}\natexlab{}.
\newblock \showarticletitle{Effectiveness of cognitive tutor algebra I at scale}.
\newblock \bibinfo{journal}{\emph{Educational Evaluation and Policy Analysis}} \bibinfo{volume}{36}, \bibinfo{number}{2} (\bibinfo{year}{2014}), \bibinfo{pages}{127--144}.
\newblock


\bibitem[Research(2020)]%
        {ixleffect2023}
\bibfield{author}{\bibinfo{person}{IXL Research}.} \bibinfo{year}{2020}\natexlab{}.
\newblock \bibinfo{title}{The IXL Effect:Measuring the Impact of IXL Math and IXL Language Arts in Pennsylvania Schools}.
\newblock
\newblock
\urldef\tempurl%
\url{https://www.ixl.com/research/Impact-of-IXL-in-Pennsylvania.pdf}
\showURL{%
\tempurl}


\bibitem[Ritter et~al\mbox{.}(2007)]%
        {ritter2007cognitive}
\bibfield{author}{\bibinfo{person}{Steven Ritter}, \bibinfo{person}{John~R Anderson}, \bibinfo{person}{Kenneth~R Koedinger}, {and} \bibinfo{person}{Albert Corbett}.} \bibinfo{year}{2007}\natexlab{}.
\newblock \showarticletitle{Cognitive Tutor: Applied research in mathematics education}.
\newblock \bibinfo{journal}{\emph{Psychonomic bulletin \& review}}  \bibinfo{volume}{14} (\bibinfo{year}{2007}), \bibinfo{pages}{249--255}.
\newblock


\bibitem[Sales and Pane(2015)]%
        {sales2015exploring}
\bibfield{author}{\bibinfo{person}{Adam~C Sales} {and} \bibinfo{person}{John~F Pane}.} \bibinfo{year}{2015}\natexlab{}.
\newblock \showarticletitle{Exploring causal mechanisms in a randomized effectiveness trial of the cognitive tutor}. In \bibinfo{booktitle}{\emph{Proceedings of the 8th International Conference on Educational Data Mining}}. \bibinfo{address}{Madrid, Spain}.
\newblock


\bibitem[Thomas et~al\mbox{.}(2023)]%
        {thomas2023ai}
\bibfield{editor}{\bibinfo{person}{Danielle~R. Thomas}, \bibinfo{person}{Jionghao Lin}, {and} \bibinfo{person}{Kenneth~R Koedinger}} (Eds.). \bibinfo{year}{2023}\natexlab{}.
\newblock \bibinfo{booktitle}{\emph{Towards the Future of AI-augmented Tutoring in Math Learning}}. \bibinfo{address}{Tokyo, Japan}.
\newblock
\urldef\tempurl%
\url{http://ceur-ws.org/Vol-3491/}
\showURL{%
\tempurl}


\bibitem[{U.S. Department of Education. Institute of Education Sciences, National Center for Education Statistics, National Assessment of Educational Progress (NAEP)}(2022)]%
        {usdoe2022naep}
\bibfield{author}{\bibinfo{person}{{U.S. Department of Education. Institute of Education Sciences, National Center for Education Statistics, National Assessment of Educational Progress (NAEP)}}.} \bibinfo{year}{2022}\natexlab{}.
\newblock \bibinfo{title}{National Assessment of Educational Progress (NAEP) Report}.
\newblock
\newblock


\bibitem[West(2023)]%
        {west2023edtech}
\bibfield{author}{\bibinfo{person}{Mark West}.} \bibinfo{year}{2023}\natexlab{}.
\newblock \bibinfo{booktitle}{\emph{An Ed-Tech Tragedy? Educational Technologies and School Closures in the time of COVID-19}}.
\newblock \bibinfo{type}{{T}echnical {R}eport}. \bibinfo{institution}{United Nations Educational, Scientific and Cultural Organization}, \bibinfo{address}{7, place de Fontenoy, 75352 Paris 07 SP, France}.
\newblock
\urldef\tempurl%
\url{https://teachertaskforce.org/sites/default/files/2023-09/2023_UNESCO_An-ed-tech-tragedy_Educational-technologies-and-school-closures-in-the-time-of-COVID19_EN_.pdf}
\showURL{%
\tempurl}


\bibitem[Zheng et~al\mbox{.}(2019)]%
        {zheng2019using}
\bibfield{author}{\bibinfo{person}{Guoguo Zheng}, \bibinfo{person}{Stephen~Edward Fancsali}, \bibinfo{person}{Steven Ritter}, {and} \bibinfo{person}{Susan Berman}.} \bibinfo{year}{2019}\natexlab{}.
\newblock \showarticletitle{Using instruction-embedded formative assessment to predict state summative test scores and achievement levels in mathematics}.
\newblock \bibinfo{journal}{\emph{Journal of Learning Analytics}} \bibinfo{volume}{6}, \bibinfo{number}{2} (\bibinfo{year}{2019}), \bibinfo{pages}{153--174}.
\newblock


\end{thebibliography}










\end{document}